\documentclass[twocolumn,showpacs,pra,aps]{revtex4-1}
\usepackage{graphicx}
\usepackage{epstopdf}
\usepackage{amsmath}
\usepackage{amssymb}
\usepackage{epsfig}
\usepackage{amsthm}
\usepackage{color}

\usepackage{graphicx,graphics,epsfig}   % Include figure files
\usepackage{dcolumn}    % Align table columns on decimal point
\usepackage{bm}         % bold math
\usepackage{amsmath}    % need for subequations
\usepackage{verbatim}   % useful for program listings
\usepackage{color}      % use if color is used in text
\usepackage{subfigure}  % use for side-by-side figures
\usepackage{times,natbib}
\usepackage{amsmath,amsfonts,amssymb,graphics,graphics,color,times}

\usepackage{latexsym}
\usepackage{amsmath}
\usepackage{amssymb}
\usepackage{amsfonts}
\usepackage{amsthm}
\usepackage{mathrsfs}
\usepackage{color,verbatim,graphics}
\usepackage[colorlinks,linkcolor=magenta,urlcolor=blue]{hyperref}  %Hyperlinks (pink, green, blue)
\usepackage{psfrag}
\DeclareMathAlphabet{\mathrsfs}{U}{rsfs}{m}{n}
\DeclareMathAlphabet{\mathpzc}{OT1}{pzc}{m}{it}
\DeclareMathAlphabet{\matheus}{U}{eus}{m}{n}
\DeclareMathAlphabet{\mathbbold}{U}{bbold}{m}{n}

\newcommand{\CC}{\mathbb{C}}

\newcommand{\ba}{\begin{eqnarray}}
\newcommand{\ea}{\end{eqnarray}}
\newcommand{\ban}{\begin{eqnarray*}}
\newcommand{\ean}{\end{eqnarray*}}
\newcommand{\Tr}{\operatorname{Tr}}%from Miguel

\newcommand{\ket}[1]{|#1\rangle}
\newcommand{\bra}[1]{\langle#1|}

\newcommand{\one}{\mathbbold{1}}

\begin{document}

\title{Measurement incompatibility does not give rise to Bell violation in general}

%\title{Notes: Triplewise incompatible measurements which cannot lead to Bell nonlocality}

\author{Erika Bene}
%\affiliation{Institute for Nuclear Research, Hungarian Academy of Sciences, H-4001 Debrecen, P.O. Box 51, Hungary}

\author{Tam\'as V\'ertesi}
\affiliation{Institute for Nuclear Research, Hungarian Academy of Sciences, H-4001 Debrecen, P.O. Box 51, Hungary}

%\affiliation{MTA Atomki, Debrecen, Hungary}

\date{\today}

%______________________________________________________________________ ABSTRACT

\begin{abstract}
In the case of a pair of two-outcome measurements incompatibility is equivalent to Bell nonlocality. Indeed, any pair of incompatible two-outcome measurements can violate the Clauser-Horne-Shimony-Holt Bell inequality, which has been proven by Wolf et al.~[Phys. Rev. Lett. 103, 230402 (2009)]. In the case of more than two measurements the equivalence between incompatibility and Bell nonlocality is still an open problem, though partial results have recently been obtained. Here we show that the equivalence breaks for a special choice of three measurements. In particular, we present a set of three incompatible two-outcome measurements, such that if Alice measures this set, independent of the set of measurements chosen by Bob and the state shared by them, the resulting statistics cannot violate any Bell inequality. On the other hand, complementing the above result, we exhibit a set of $N$ measurements for any $N>2$ that is $(N-1)$-wise compatible, nevertheless it gives rise to Bell violation.
\end{abstract}

\maketitle

\section{Introduction}

Correlations resulting from incompatible local measurements on an entangled quantum state can violate Bell inequalities~\cite{bell,brunnerreview}. However, Bell violation is not possible if either the measurements are compatible or the shared state is unentangled. In this respect, one may ask whether (i) all entangled states lead to Bell violation. This turns out not to be true for projective measurements~\cite{werner} and for the general case of positive-operator-valued-measure (POVM) measurements as well~\cite{barrett} (see also Refs.~\cite{betterhirsch,simpovm} for more recent results). Similarly, one may ask whether (ii) all incompatible measurements lead to Bell violation. Specifically, the question is whether for any given set of incompatible measurements performed by Alice, one can always find a shared entangled state and a set of measurements for Bob, such that the resulting statistics will lead to Bell inequality violation.

This holds true in the case of any number of incompatible projective measurements~\cite{khalfin}, and for a pair of dichotomic measurements as well~\cite{wolf}. However, in the case of more than two non-projective dichotomic measurements (or in the case of two non-dichotomic measurements) the problem is still open. Though, there is recent progress toward this aim. For example, a strong link between incompatibility of measurements and Einsten-Podolsky-Rosen (EPR) steering~\cite{EPR1,EPR2}, a phenomenon in between entanglement and Bell nonlocality, has been established~\cite{uola14,q14,q16}.

In this paper, we present a set of three incompatible dichotomic measurements, such that if Alice uses this triple, independent of the set of measurements chosen by Bob and the state shared by them, the resulting statistics cannot violate any Bell inequality. This result remains valid for Bell inequalities with arbitrary number of settings and outcomes on Bob's side, including the general case that Bob is allowed to carry out arbitrary POVM measurements. Note that the case where Bob's settings are restricted to projective measurements have been settled recently~\cite{q16}.

In addition, and complementary to the above results, we present a set of $N$ measurements, such that any $N-1$ measurements out of this set are compatible. However, we show that using this set of $N$ measurements on one side, and another set of $N$ measurements on the other side along with a suitable shared state between them leads to violation of a Bell inequality. This result holds true for any number of $N>2$ settings.

The paper is structured as follows. In Sec.~\ref{setup}, we start by defining the setup and we fix notation. Sec.~\ref{proof} is devoted to the detailed proof of our main result. To do so, we simplify the problem in Sec.~\ref{simplify} by showing that given Alice's specific set of three measurements, it is sufficient to deal with pure two-qubit states in the Schmidt form $\ket{\psi} = \cos\theta\ket{00}+\sin\theta\ket{11}$ along with Bob's real-valued ternary-outcome POVMs. Then, depending on the value of the parameter $\theta$, we will split the proof into two parts. The case of small $\theta\le\theta^*$ values are considered in Sec.~\ref{proofa}, whereas the case of large $\theta>\theta^*$ values are treated in Sec.~\ref{proofb}. Then in Sec.~\ref{compl}, complementing the above results, we exhibit $N$ measurements for any $N>2$ that are $(N-1)$-wise compatible, however they give rise to Bell violation. The paper ends with conclusion in Sec.~\ref{conclusion}.

\section{Setup}
\label{setup}

\begin{figure}[h]
\centering
\includegraphics[width=0.9\columnwidth,trim={0.5cm 0.0cm 0.0cm 0.5cm},clip]{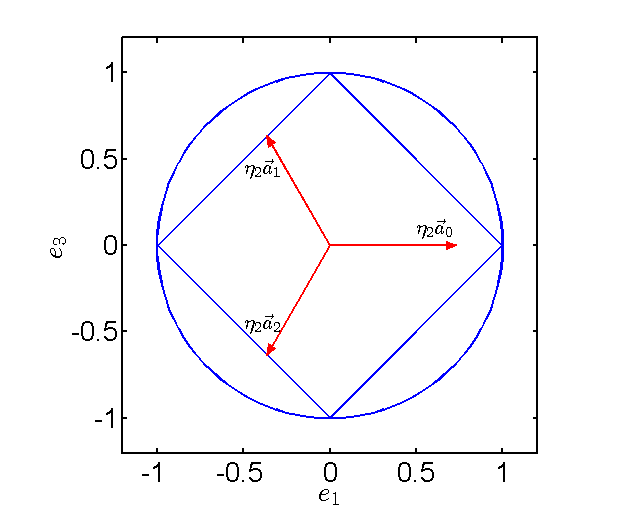}
\caption{Any vector within the square area can be decomposed from the four Bloch vectors $\pm\vec e_1$ and $\pm\vec e_3$ pointing toward the corners of the square inscribed in the unit circle. These corners correspond to the respective Pauli measurements $\{P_{a|0}, a=0,1\}$ and $\{P_{a|1}, a=0,1\}$. The shrunk Bloch vectors $\eta_2\vec a_x$, $x=0,1,2$ with $\eta_2=\sqrt3-1$  define the noisy trine measurements, which are lying within the square, hence simulable with the two Pauli measurements above.}
\label{fig1}
\end{figure}

A general quantum measurement is represented by a set of positive definite operators $\{M_a\}$, $M_a\ge 0$ that sum to the identity, $\sum_{a}M_a = \one$. We consider the following set of three dichotomic qubit POVMs, so-called trine measurements (labeled by $x=0,1,2$):
\begin{equation}
\label{trine1}
M^{\eta}_{a|x}=\frac{1}{2}\left(\one+(-1)^a\eta\vec a_x\cdot\vec\sigma\right),
\end{equation}
where $a$ labels the two possible outcomes $\{0,1\}$, and the vector $\vec\sigma=(\sigma_1,\sigma_2,\sigma_3)$ stand for the three Pauli matrices $X$, $Y$, and $Z$, respectively. Above $\eta$ is a parameter between zero and one. In case of $\eta=1$, the measurement is projective, and in case of $\eta=0$, the measurement is the identity.
The three Bloch vectors of Alice's measurements are chosen as
\begin{equation}
\label{trine2}
\vec a_x = \cos(2x\pi/3)\vec e_1 + \sin(2x\pi/3)\vec e_3
\end{equation}
for $x=0,1,2$. That is, the three measurement directions $\vec a_x$, $(x=0,1,2)$ point toward the vertices of a regular triangle on the real plane (see Fig.~\ref{fig1}).

Let us now define what we mean by incompatibility of a given set of $n$ measurements. We say that Alice's set of measurements $\{M_{a|x}\}$, $x=(1,\ldots,n)$ is $n$-wise jointly measurable~\cite{busch,buschbook}, if there exists a $2^n$-outcome parent measurement with POVM elements $M_{\textbf a}$, such that each outcome corresponds to a bit string $\textbf a=(a_1,a_2,\ldots,a_n)$ such that
\begin{equation}
\label{parent}
M_{a_x|x}=\sum_{\textbf a\setminus a_x}M_{\textbf a},
\end{equation}
where the notation $\textbf a\setminus a_x$ stands for an $(n-1)$ bit string formed of all the bits of $\textbf a$ except for $a_x$.

If the set $\{M_{a|x}\}$ is not $n$-wise jointly measurable, the set is said to be incompatible. Specifically, the measurements given by Eqs.~(\ref{trine1},\ref{trine2}) are known to be pairwise jointly measurable below $\eta_2 = \sqrt 3 - 1\simeq 0.7321$ and triplewise jointly measurable below $\eta_3 = 2/3$~\cite{liang,hrs,uola16}. Hence, there is a range $2/3<\eta<\sqrt3-1$, where the set forms a so-called hollow triangle~\cite{q14}: In this range, the set of three POVMs is pairwise jointly measurable, but not triplewise jointly measurable, hence the three measurements are incompatible.

Let us now fix $\eta^*=0.67$. According to the above, the set $\{M_{a|x}^{\eta^*}\}$ defines a hollow triangle. In this notes, we show that there is no Bell inequality which can be violated if Alice measures this set. Namely, we show that the probability distribution
\begin{equation}
\label{stat}
p(ab|xy)=\Tr(\rho M_{a|x}^{\eta^*}\otimes M_{b|y}),\,x=0,1,2,\, a=0,1
\end{equation}
is local for any state $\rho$ shared by Alice and Bob and arbitrary measurements~$\{M_{b|y}\}$ (including an arbitrary number of settings $y$ and outcomes $b$ for Bob). Note that a probability distribution $p(ab|xy)$ is local if and only if it admits a decomposition of the form
\begin{equation}
\label{LHV}
p(ab|xy)=\sum_{\lambda}{p(\lambda)p_A(a|x,\lambda)p_B(b|y,\lambda)},
\end{equation}
where $\lambda$ is a shared variable and $p(\lambda)$ defines weights summing up to 1, whereas $p_A$ and $p_B$ define Alice and Bob's respective local response functions. The construction of such a local hidden variable (LHV) model will prove our assertion that measurement incompatibility does not imply Bell nonlocality in general. Below we present the detailed proof, which starts with a slight simplification of the problem.

\section{Proof}
\label{proof}

\subsection{Simplification}
\label{simplify}
First, instead of a general mixed state $\rho$ in Eq.~(\ref{stat}) we can consider pure states without loss of generality~\cite{q16}. This is due to the convexity of the set of local correlations and the fact that $\rho$ depends linearly on the probabilities $p(ab|xy)$ in Eq.~(\ref{LHV}). Next, since Alice's measurements~(\ref{trine1}) act on a qubit, the shared state takes the general form of two-qubit pure states
\begin{equation}
\label{psi}
\ket{\psi}=U_A\otimes U_B(\cos\theta\ket{00} + \sin\theta\ket{11}),
\end{equation}
where $\theta\in[0,\pi/4]$ and $U_A, U_B$ are arbitrary (unitary) qubit rotations. On the other hand, Bob's set of measurements
$\{M_{b|y}\}$ are qubit POVMs (with possibly infinite number of inputs $y$ and outputs $b$). Furthermore, instead of generic qubit $U_A$ and $U_B$ unitaries we can choose $U_A=O(\varphi)$ and
$U_B=\one$ in the state~(\ref{psi}), where $O(\varphi)$ is given
by a planar rotation
\begin{equation}
\label{Orot}
O(\varphi)=\cos\varphi(\ket{0}\bra{0}+\ket{1}\bra{1})+\sin\varphi(-\ket{0}\bra{1}+\ket{1}\bra{0})
\end{equation}
and we can further assume that Bob's measurements $\{M_{b|y}\}$ are real valued. The corresponding proofs are deferred to Appendix~\ref{appnew}. In addition, since any extremal real-valued qubit POVM has at most three outcomes~\cite{dariano05}, this entails that it suffices to consider Bob's real-valued measurements with at most three outcomes (that is, $b\in\{0,1,2\})$.

Due to the above simplifications, the proof boils down to show that the probability distribution
\begin{align}
\label{pQtp}
p(ab|xy)=&\Tr{(\rho(\theta,\varphi)M_{a|x}^{\eta^*}\otimes M_{b|y})},\nonumber\\
&x=0,1,2,\, a=0,1,
\end{align}
where $\eta^* = 0.67$, admits a LHV model in the form~(\ref{LHV}), where the two-parameter family of states
\begin{equation}
\label{rhotp}
\rho(\theta,\varphi)=\ket{\psi(\theta,\varphi)}\bra{\psi(\theta,\varphi)}
\end{equation}
is as follows
\begin{equation}
\label{psitp}
\ket{\psi(\theta,\varphi)}=O_A(\varphi)\otimes\one(\cos\theta\ket{00}+\sin\theta\ket{11}),
\end{equation}
and the set $\{M_{b|y}\}$ consists of an arbitrary number of real valued qubit measurements $y$ with ternary outcomes $b=0,1,2$.

As we stated in the introduction, the proof will be split into two parts, the case of small values ($\theta<\theta^*$), and the case of large values
($\theta^*<\theta\le\pi/4$), where the threshold $\theta^*$ appears to be
\begin{equation}
\label{thetastar}
\theta^*=\frac{1}{2}\arcsin\sqrt{\left(\frac{100}{67}\right)^2\left(\sqrt 3 -1\right)^2-1}\simeq0.2279.
\end{equation}

Let us first start with the case of small $\theta$ values.

\subsection{Small $\theta$ values}
\label{proofa}

In this regime the proof is fully analytical. Let us consider the two Pauli measurements $\sigma_1=X$ and $\sigma_3=Z$ with respective projectors
\begin{align}
\label{paulixz}
P_{a|0}=(\one+(-1)^a X)/2,\nonumber\\
P_{a|1}=(\one+(-1)^a Z)/2,
\end{align}
where $a\in\{0,1\}$. We next consider the noisy trine measurements defined by the formulas~(\ref{trine1},\ref{trine2}), where the three shrunk vectors $\eta\vec a_x$, $(x=0,1,2)$ point toward the vertices of an equilateral triangle (see Fig.~\ref{fig1}). It is a simple exercise to show that the shrunk vectors are inside the square spanned by the unit vectors $\pm\vec e_1$ and $\pm\vec e_3$ if
\begin{equation}
\label{etacs}
\eta\le\eta_2=\sqrt3-1\simeq 0.7321.
\end{equation}
Therefore the noisy trine measurements~(\ref{trine1},\ref{trine2}) for $\eta\le\eta_2$ can be expressed as convex combinations of the two Pauli measurements $X$ and $Z$. In other words, given an input choice (one of the noisy trine measurements), one can translate it into choosing one of the two Pauli measurements $X$ and $Z$ along with some randomness~\cite{finitesimulate}.

%Then, it follows from the argument of Ref.~\cite{finitesimulate} that the noisy trine measurements~(\ref{trine1},\ref{trine2}) for $\eta\le\eta_2$ can be classically simulated by the two Paulis $X$ and $Z$.

Similarly, if we have noisy Pauli measurements
\begin{align}
\label{tildev}
P^v_{a|0}=(\one+(-1)^a v X)/2,\nonumber\\
P^v_{a|1}=(\one+(-1)^a v Z)/2,
\end{align}
where $a\in\{0,1\}$, the trine measurements~(\ref{trine1},\ref{trine2}) can be simulated up to a visibility of $\eta=v\eta_2$ with measurements~$(\ref{tildev})$.

Suppose now that the distribution
\begin{align}
\label{lhv1}
p(ab|xy)=&\Tr{\left(\rho(\theta,\varphi)P^v_{a|x}\otimes M_{b|y}\right)}\nonumber\\
&x=0,1,\, a=0,1,
\end{align}
admits a LHV model for some $v$, where the state $\rho(\theta,\varphi)$ is defined by Eqs.~(\ref{rhotp},\ref{psitp}), $P^v_{a|x}$ by Eq.~(\ref{lhv1}), and $\{M_{b|y}\}$ is an arbitrary set of qubit measurements on Bob's side. Then the simulability of the trine measurements with the noisy Paulis $(\ref{tildev})$ above entails that the distribution
\begin{equation}
\label{lhv2}
p(ab|xy)=\Tr{\left(\rho(\theta,\varphi)M^{v\eta_2}_{a|x}\otimes M_{b|y}\right)},
\end{equation}
admits a LHV model as well, where $M_{a|x}^{v\eta_2}$ are the trine measurements~(\ref{trine1},\ref{trine2}) with a visibility of $v\eta_2=v(\sqrt 3 -1)$. Indeed, if the distribution $p(ab|xy)$ in Eq.~(\ref{lhv2}) was nonlocal, i.e. there existed a Bell inequality violated by $p(ab|xy)$, the use of measurements~(\ref{tildev}) in Eq.~(\ref{lhv2}) would give at least the same Bell violation due to the above simulability results of measurements and the linearity of the trace rule. This is a contradiction, hence the distribution~(\ref{lhv2}) has to admit a LHV model.

Let us now invoke Ref.~\cite{pironio}, where it has been proven that the Clauser-Horne-Shimony-Holt (CHSH) inequality~\cite{chsh} is the only inequivalent Bell inequality in the bipartite scenario, where Alice has two dichotomic settings and Bob has any number of settings $y$ with arbitrary number of outcomes $b$. Therefore, a probability distribution $p(ab|xy)$ where $a,x=0,1$, and $b,y$ are possibly infinite, admits a LHV model if and only if $p(ab|xy)$ does not violate (any of the versions of) the CHSH inequality. Put together with the above simulability result, if the probability distribution~(\ref{lhv1}) does not give rise to Bell-CHSH-violation, it implies that the probability distribution~(\ref{lhv2}) admits a LHV model.

Then it is enough to check the range of parameters ($\theta,\varphi,v$) for which the distribution~(\ref{lhv1}) does not give rise to CHSH violation. Due to the Horodecki criterion~\cite{horo}, a pure two-qubit state~(\ref{psitp}) has a maximal CHSH violation of $2\sqrt{1+\sin^2{2\theta}}$, which value can be attained with the Pauli measurements~(\ref{paulixz}) (in some rotated bases on Alice's side). Note that this violation is independent of the angle $\varphi$. Also, for the noisy Paulis~(\ref{tildev}) with visibility $v$, the maximum CHSH value becomes $2v\sqrt{1+\sin^2{2\theta}}$. Since the local bound of the CHSH inequality is 2, we get the criterion
\begin{equation}
\label{vcs} v\le v^*=\frac{1}{\sqrt{1+\sin^2{2\theta}}}.
\end{equation}
to have a local model for the distribution~(\ref{lhv1}) using a two-qubit pure state~(\ref{psitp}) independently of the set of measurements chosen by Bob.

Putting all the above results together, the trine measurements~(\ref{trine1},\ref{trine2}) with a visibility of $\eta = v^*\eta_2$, where the state is defined by~(\ref{psitp}) and Bob has arbitrary measurements, gives a local distribution~$p(ab|xy)$. Above, $v^*$ is given by (\ref{vcs}) and $\eta_2$ is given by (\ref{etacs}). Suppose, we want a LHV model for $\eta=\eta^*=0.67$, then the critical $\theta^*$ below which the distribution $p(ab|xy)$ is local is given by the solution of the equation $\eta^*=67/100=v^*\eta_2$. This value is $\theta^*\simeq 0.2279\;\textrm{[rad]}$, and the exact value is given by formula~(\ref{thetastar}).

\subsection{Large $\theta$ values}
\label{proofb}

For the region $\theta^*\le\theta\le\pi/4$ we use a different approach. Recall that our task is to show that the probability distribution~(\ref{pQtp}) with $\eta^*=0.67$ admits a LHV model~(\ref{LHV}). The pure state $\rho(\theta,\varphi)$ is defined by Eq.~(\ref{psitp}), where we now focus on the range $\theta^*<\theta\le\pi/4$ and $0\le\varphi\le2\pi$, where $\theta^*$ is given by Eq.~(\ref{thetastar}). On the other hand, Bob's set of measurements $\{M_{b|y}\}$ consists of an arbitrary number of real valued qubit measurements $y$ with ternary outcomes each (that is we have $b\in\{0,1,2$\} for each setting $y$). Our procedure is based on discretizing the set $\theta\in[\theta^*,\pi/4]$. Note that a similar procedure has been carried out in Refs.~\cite{uola14,hirsch16}.

In particular, we give a linear program in Sec.~\ref{fixangles} which lowerbounds the value of $\eta$ considering any fixed state $\rho(\theta,\varphi)$ in Eqs.~(\ref{rhotp},\ref{psitp}), for which a LHV model exists. Defining a fine enough grid for $\theta$ and $\varphi$, and taking the minimum $\eta$ over the grid points allow us to lowerbound $\eta$ globally for this particular grid. Then, in Sec.~\ref{allangles} the continuous case will be considered. In particular, starting from a finite set $\{(\theta_i,\varphi_i), i=1,\ldots,n\}$, which gives us a LHV model for $\eta(\theta_i,\varphi_i)$, we provide a LHV model for $\eta=0.67$ for a continuous values of $(\theta,\varphi)$. The treatment of this continuous case is based on the method presented in Ref.~\cite{q16}.

\subsubsection{Finite grid}
\label{fixangles}

In order to lowerbound $\eta$ for any given pair of angles $(\theta,\varphi)$, we first discretize Bob's POVM measurements using the method presented in Ref.~\cite{alg1} (see Appendix~A of this reference for the case of general POVM measurements). Instead of considering an infinite continuous set, we take a finite number of POVM elements $\{M_{b|y}\}$. Given this finite set of POVM elements, one can simulate a continuous set of (noisy) measurements for some $\eta_B$
\begin{equation}
\label{Mb}
M_{b}^{\eta_B}=\eta_B M_{b}+(1-\eta_B)\Tr{(M_{b}\zeta_B)}\one,
\end{equation}
where $\{M_b, (b=0,1,2)\}$ is an arbitrary three-outcome POVM on the real plane, and $\zeta_B$ is some fixed qubit state. The above simulation means that $M_{b}^{\eta_B}$ can always be written as a convex combination of the finite number of POVM elements $\{M_{b|y}\}$. In particular, we pick a finite set consisting of 9 binary-outcome and 4 ternary-outcome measurements. The binary-outcome measurements
\begin{equation}
\label{Mbin}
M_{b|y}=\frac{\one+(-1)^b\vec u_y\cdot\vec\sigma}{2},\quad b=0,1
\end{equation}
are defined by the Bloch vectors
\begin{equation}
\vec u_y = \cos(y\pi/9)\vec e_1 + \sin(y\pi/9)\vec e_3,
\end{equation}
where $y=(0,1,\ldots,8)$. On the other hand, the ternary-outcome measurements $M_{b|y}$, $y=(9,10,11,12)$ are defined by the three POVM elements as follows
\begin{align}
\label{Mternary}
M_{0|y}&=(\one+\vec v_{0y}\cdot\vec\sigma)/3,\nonumber\\
M_{1|y}&=(\one+\vec v_{1y}\cdot\vec\sigma)/3,\nonumber\\
M_{2|y}&=(\one+\vec v_{2y}\cdot\vec\sigma)/3,
\end{align}
where the respective Bloch vectors are
\begin{align}
\vec v_{0y}&=\cos(y\pi/2)\vec e_1+\sin(y\pi/2)\vec e_3,\nonumber\\
\vec v_{1y}&=\cos(y\pi/2+2\pi/3)\vec e_1 + \sin(y\pi/2+2\pi/3)\vec e_3,\nonumber\\
\vec v_{2y}&=\cos(y\pi/2+4\pi/3)\vec e_1 + \sin(y\pi/2+4\pi/3)\vec e_3
\end{align}
for $y=9,10,11$, and $12$. In addition, we also include the three degenerate measurements and the six different outcome relabellings of each POVM $M_{b|y}$, $b=0,1,2$, for all $y=0,1,\ldots,12$ in the finite set, where the binary-outcome measurements are embedded into the space of three-outcome POVM elements. This amounts to $3+6\times13=81$ POVMs, which define a polytope with 81 vertices, whose facets can be determined using a polytope software. Let us define $\zeta_B$ through $\alpha$ as follows
\begin{equation}
\label{zetaB}
\zeta_B=\alpha\ket{0}\bra{0}+(1-\alpha)\one/2.
\end{equation}
We choose two distinct $\alpha$ values, $\alpha=0$ and $\alpha=5/6$. Following the method in the Appendix of Ref.~\cite{alg1} and running the program cdd~\cite{cdd}, we get the threshold values $\eta_B=0.9268$ for $\alpha=0$ and $\eta_B=0.8900$ for $\alpha=5/6$. Therefore, we can express Bob's (noisy) measurements $M_{b}^{\eta_B}$ in Eq.~(\ref{Mb}) by the above $\eta_B$ values as a convex combination of the 81 POVMs above.

We are now ready to use the trick of Refs.~\cite{alg1,alg2} to simulate a distribution $p(ab|xy)$ coming from a continuous set of Bob's measurements $M_{b}$ using a finite set $\{M_{b|y}\}$. The optimization problem below is a modified version of Protocol~2 in \cite{alg1}:
\begin{equation} \label{linprog}
\begin{aligned}
\text{max}\quad & \eta \\
\text{subject to}\quad &\Tr{(M_{a|x}^{\eta}\otimes M_{b|y}\chi)}=\sum_{\lambda}{p_{\lambda}D_{\lambda}}\\
\quad &\sum_{\lambda}{p_{\lambda}}=1, p_{\lambda}\ge 0\quad \forall \lambda,\quad\forall a,b,x,y\\
\quad &\rho(\theta,\varphi) = \eta_B\chi + (1-\eta_B)\chi_A\otimes\zeta_B
\end{aligned}
\end{equation}
The input to this program are $\eta_B$ and $M_{b|y}$ from Eqs.~(\ref{Mbin},\ref{Mternary}), $\rho(\theta,\varphi)$ in Eq.~(\ref{rhotp}), and the deterministic strategies $D_{\lambda}$. On the other hand, the optimization variables are $p_{\lambda}$ and $\chi$. This is not a linear program yet, however notice that $\chi_A=\rho_A\equiv\Tr_B{\rho(\theta,\varphi)}$ and $\chi$ can be expressed from the last line of the problem~(\ref{linprog}) as
\begin{equation}
\label{chi}
\chi=\frac{1}{\eta_B}\rho(\theta,\varphi)+\frac{\eta_B-1}{\eta_B}\rho_A\otimes\zeta_B.
\end{equation}
This allows us to obtain the following linear program:
\begin{equation} \label{linprog2}
\begin{aligned}
\text{max}\quad & \eta \\
\text{subject to}\quad &\Tr{(M_{a|x}^{\eta}\otimes M_{b|y}\chi)}=\sum_{\lambda}{p_{\lambda}D_{\lambda}}\\
\quad &\sum_{\lambda}{p_{\lambda}}=1, p_{\lambda}\ge 0 \quad \forall \lambda,\quad\forall a,b,x,y,
\end{aligned}
\end{equation}
where the input $\chi$ and $M_{b|y}$ come from Eq.~(\ref{chi}) and Eqs.~(\ref{Mbin},\ref{Mternary}), respectively, and the optimization variables are $p_{\lambda}$. Note that we can further write $\rho_A=O(\varphi)\Tr_B{\rho(\theta,0)}O(\varphi)$.

Calling the solver Mosek~\cite{mosek} either with $\alpha=0$ or $\alpha=5/6$, it takes about 7 sec to solve the linear program~(\ref{linprog}) and return $\eta$ in our standard desktop PC for a fixed value of $\rho(\theta,\varphi)$. Let us denote $\bar{\eta} = \max\{\eta(\alpha=0),\eta(\alpha=5/6)\}$ for a given pair $(\theta,\varphi)$. The above program allows us to evaluate $\bar{\eta}$ for any fixed $(\theta,\varphi)$. Our goal is to prove that $\bar{\eta}$ is above the threshold $\eta=0.67$ in the whole interval $\theta^*\le\theta\le\pi/4$ and $0\le\varphi\le2\pi$. We cover this continuous case in the next subsection. To this end, we resort to the technique proposed in Ref.~\cite{q16}.

%As a test case, we picked $10^5$ random pair of angles in the range ($\theta^*\le\theta\le\pi/4$ and $0\le\varphi\le2\pi$), maximized $\eta$ in program~(\ref{linprog2}) both for $\alpha=0$ and $\alpha=5/6$, and picked the larger one. As a result we always found $\eta>0.67$. This supports that the threshold is above $\eta=0.67$. However, it does not cover the most general continuous case, which is treated in the next subsection.

\subsubsection{Continuous case}
\label{allangles}

We first minimized $\bar{\eta}$ in the two variables $\theta^*\le\theta\le\pi/4$ and $0\le\varphi\le2\pi$ using the heuristic search Amoeba~\cite{amoeba}, and obtained the minimum $\bar{\eta}=0.6808$ by the variables $\theta=\pi/4$, $\varphi\simeq0.1192$ and $\alpha=0$. This gives a strong numerical evidence that $\eta\ge0.67$ for the continuous case as well.

We next prove this result in a semi-analytical way. To this end, we closely follow the method introduced in Ref.~\cite{q16}. Suppose we have a state~$\rho(\theta,\varphi)$ in Eq.~(\ref{rhotp}) for $\theta=\theta_i$, $\varphi=\varphi_j$, and $\eta$ in Alice's measurements~(\ref{trine1}), such that the distribution~(\ref{pQtp}) admits a LHV model. Then we also have a LHV model for a state (with the same measurements of Alice) which is a convex mixture of our state $\rho$ and a separable state
\begin{equation}
p\rho(\theta_i,\varphi_j)+(1-p)\sigma,
\end{equation}
where $0\le p\le 1$ and $\sigma$ denotes a separable state. Let $\rho_B=\Tr_A{\rho(\theta,\varphi)}$. Therefore, if we can write
\begin{equation}
\label{pvsigma}
v\rho(\theta,\varphi)+(1-v)\frac{\one}{2}\otimes\rho_B=p\rho(\theta_i,\varphi_j) + (1-p)\sigma
\end{equation}
for some weight $p$ and separable state $\sigma$, then the distribution~(\ref{pQtp}) admits a LHV model for the state $\rho(\theta,\varphi)$ and for Alice's trine measurements $M_{a|x}^{v\eta}$ in Eq.~(\ref{trine1}). Let us note that in order to get the above equation, we also passed an amount of $(1-v)$ noise from Alice's measurements to the state. We expect to find such a decomposition in~(\ref{pvsigma}) which in the limits $\theta\rightarrow\theta_i$, $\varphi\rightarrow\varphi_j$ gives us the value of $v$ close to 1. Recall that we obtained $\eta$ larger than 0.6808 over all $(\theta,\varphi)$ using a heuristic search. Hence, if we can make a fine enough grid of the $(\theta_i,\varphi_j)$ values with $\eta\ge0.6808$ for all grid points, we expect to have $\eta(\theta,\varphi)=v\eta(\theta_i,\varphi_j)>0.67$ for the continuous case $(\theta,\varphi)$. Note also that due to symmetries it is enough to consider the regime $\varphi\in\{0,\pi/6\}$ and $\theta\in\{0.2279,\pi/4\}$.

We have to discuss two separate cases according to the movement from the coordinate~($\theta_i,\varphi_j$) to the two orthogonal directions. In the case of both directions, we start from a pair $(\theta_i,\varphi_j)$ and a fixed $\alpha$, either $0$ or $5/6$, and call the linear program~(\ref{linprog2}) to compute $\eta$. Then we find analytical formulas which allow us to obtain $\eta(\theta,\varphi_j)$ in the case of $\theta = \theta_i - \delta\theta$, and $\eta(\theta_i,\varphi)$ in the case of $\varphi = \varphi_j + \delta\varphi$. The respective formulas are as follows:
\begin{equation}
\label{etaform1}
\eta(\theta,\varphi_j) = \frac{\eta(\theta_i,\varphi_j)}{\cot\theta\tan\theta_i(1+\eta(\theta_i,\varphi_j))-\eta(\theta_i,\varphi_j)}
\end{equation}
and
\begin{equation}
\label{etaform2}
\eta(\theta_i,\varphi) = \frac{1-2\sqrt 2\sqrt{1-\cos(2\delta\varphi)}}{8\cos(2\delta\varphi)-7}\eta(\theta_i,\varphi_j),
\end{equation}
where the proofs are given in Appendices~\ref{appB} and~\ref{appC}. These formulas give us a method to tackle the continuous case $(\theta,\varphi)$ given the values of $\eta$ for a finite grid $\{(\theta_i,\varphi_j)\}$.

\begin{figure}[h]
\centering
\includegraphics[width=1.05\columnwidth,trim={1.5cm 0.0cm 2.0cm 0.0cm},clip]{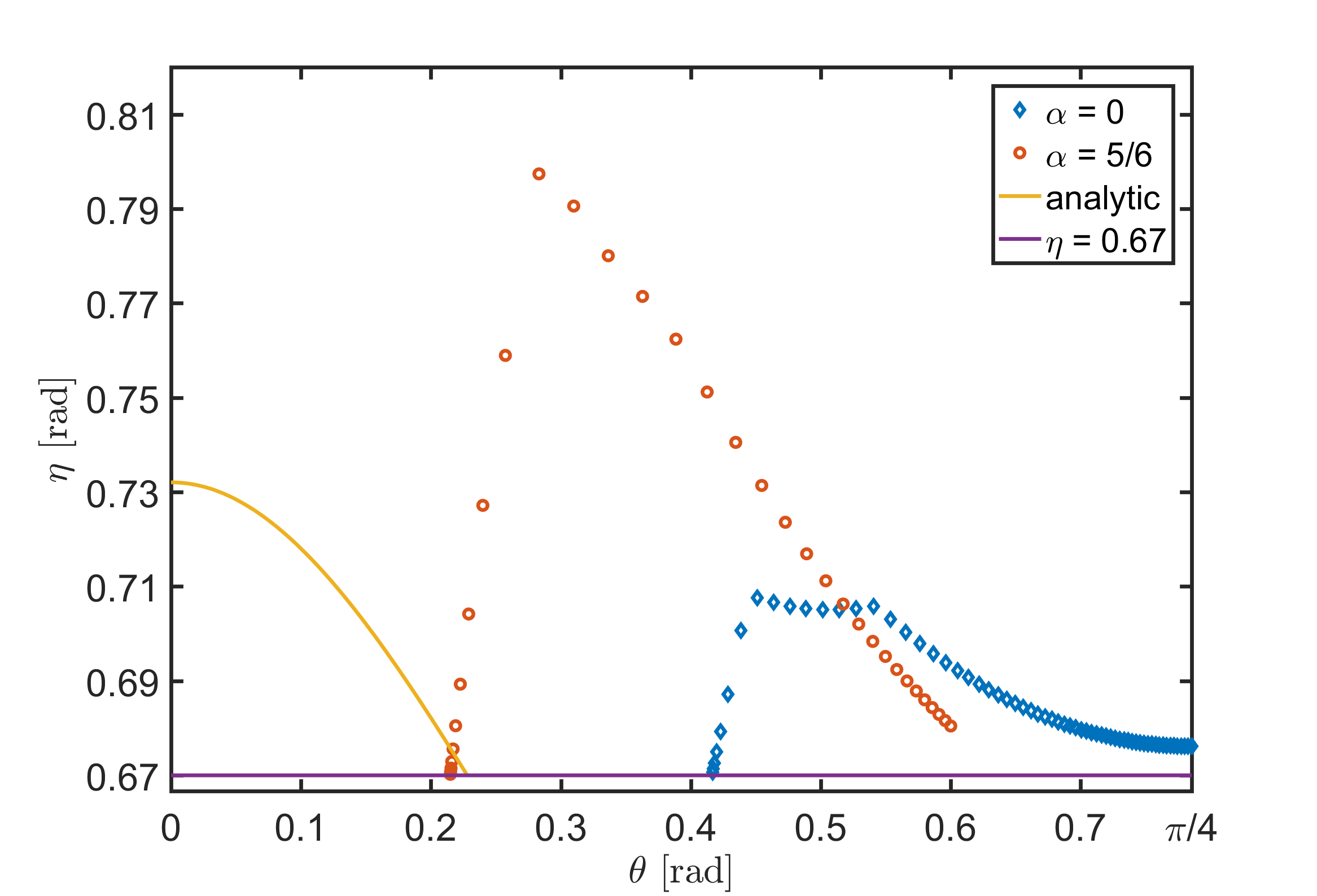}
\caption{The graphs demonstrate that $\eta$ is larger than $0.67$ for any $0\le\theta\le\pi/4$ and any $0\le\varphi\le 2\pi$. The solid curve shows the small $\theta$ region, where the proof is analytic. The markers show the semi-analytic region. The diamond markers denote $\alpha=0$, whereas open circles designate the case $\alpha=5/6$.}
\label{fig2}
\end{figure}

Given these formulas, we first find a lower bound on $\eta(\theta_i)\equiv\min_{\varphi}\eta(\theta_i,\varphi)$ for a fixed $\theta_i$ value, where optimization is carried out over all $\varphi$. We use Eq.~(\ref{etaform2}) and set $\delta\varphi=\varphi-\varphi_j=0.1$ degree to obtain a lower bound of $\eta(\theta_i,\varphi)=v\eta(\theta_i,\varphi_j)$, where $v=0.993067$. Therefore, in order to get a lower bound for a given angle $\theta_i$ and all $\varphi$ we have to compute
\begin{equation}
\label{etaAphi}
\eta(\theta_i)=0.993067\times\min_j\{\eta(\theta_i,\varphi_j)\},
\end{equation}
where the angles $\varphi_j$ scan the discrete set $\varphi_j=[0,0.1,0.2,\ldots,29.8,29.9,30]$ degrees consisting of
301 different angles. This method provides us with the bound $\eta(\theta_i)$ valid for a fixed $\theta_i$ and any values of $\varphi$. Note that it takes 7 sec for our computer to solve the linear program for $\eta$ in a single instance of $(\theta_i,\varphi_j)$, hence the overall time to compute $\eta(\theta_i)$ is $301\times 7$ sec, that is, roughly half an hour.

Having the above bound $\eta(\theta_i)$ for a fixed $\theta_i$, we can compute the lower bound $\eta(\theta)$ for any $0<\theta<\theta_i$ by using the formula:
\begin{equation}
\label{etaglobal}
\eta(\theta) = \frac{\eta(\theta_i)}{\cot\theta\tan\theta_i(1+\eta(\theta_i))-\eta(\theta_i)}.
\end{equation}
In this way, we get $\eta$ valid for a continuous set of $\theta$ and $\varphi$ values. The proof of the above formula is based on the fact that formula~(\ref{etaform1}) for any fixed $0<\theta<\theta_i\le\pi/4$ is a monotonic (increasing) function of $\eta$. Then, for any fixed $0<\theta<\theta_i$, we have
\begin{equation}
\min_{\varphi}{\eta(\theta,\varphi)}\ge \min_{\varphi}\frac{\eta(\theta_i,\varphi)}{\cot\theta\tan\theta_i(1+\eta(\theta_i,\varphi))-\eta(\theta_i,\varphi)},
\end{equation}
which is further lowerbounded by Eq.~(\ref{etaglobal}) due to the above mentioned monotonic property.

The actual numerical treatment for $\alpha=0$ in Eq.~(\ref{zetaB}) proceeds as follows:
\begin{enumerate}
\item Set $i=0$ and $\theta_0=\pi/4$.
\item Compute $\eta(\theta_i)$ in Eq.~(\ref{etaAphi}).
\item Compute $\theta<\theta_i$ for which $\eta(\theta)=0.67$ using formula~(\ref{etaglobal}) and identify $\theta_{i+1}=\theta$.
\item Set $i=i+1$ and go back to step 2 while $(\theta_i-\theta_{i+1})>\epsilon$, where $\epsilon$ is a small number, say, $10^{-4}$.
\end{enumerate}
We do the same computation by choosing $\alpha=5/6$ and $\theta_0=0.6$ in the first step of the algorithm above. The results are visualized in Fig.~\ref{fig2} (the diamonds stand for $\alpha=0$ and the empty circles are for $\alpha=5/6$). Let us stress that in the region in between two consecutive markers the above analytical lower bounds guarantee that $\eta$ cannot drop below $0.67$. On the other hand, the solid curve corresponds to the analytical lower bound. As we see, the three curves cover all the range $0\le\theta\le\pi/4$, which completes the proof.

\section{Bell violation with $(N-1)$-wise compatible measurements}
\label{compl}

In this section, we further explore the link between joint measurability and Bell violations. It has been proven in Ref.~\cite{q14} that there exists a specific pairwise jointly measurable set of $N=3$ dichotomic POVMs which give rise to the violation of the $I_{3322}$ three-setting two-outcome Bell inequality~\cite{I3322}. Below we generalize this result to any $N>3$. In particular, we present $N$ observables, which are $(N-1)$-wise jointly measurable, and give rise to violation of an $N$-setting Bell inequality.

To this end, we use the construction from Ref.~\cite{closing}. Namely, it has been proven there that there exist a pure quantum state $\rho$ acting on $\CC^N\otimes\CC^N$ and specific two-outcome projective measurements $\bar M_{a|x}$ and $\bar M_{b|y}$, $a,b=0,1$, $x,y=1,\ldots,N$ (defined by Eqs.~3, 4, and 5 in Ref.~\cite{closing}), giving rise to the probability distribution
\begin{align}
\label{ppp}
p(00|xy)&=\eta\Tr{\left(\rho \bar M_{0|x}\otimes \bar M_{0|y}\right)},\nonumber\\
p_A(0|x)&=\eta\Tr{\left(\rho \bar M_{0|x}\otimes \one\right)},\nonumber\\
p_B(0|y)&=\Tr{\left(\rho \one\otimes \bar M_{0|y}\right)},
\end{align}
which has been shown to violate the $N$-setting $I_{NN22}$ inequality for the parameter range $\eta\ge 1/(N-1)$. Note that we switched Alice and Bob with respect to the notation in Ref.~\cite{closing}. It is also noted that $p_A(0|x)=\sum_{y=1}^N p(0b|xy)$ and $p_B(0|y)=\sum_{x=1}^N p(a0|xy)$ stand for Alice's and Bob's respective marginal distributions. The $N$-setting Bell inequality $I_{NN22}$ was discovered by Collins and Gisin~\cite{I3322}, for which the $I_{3322}$ inequality is the first member $N=3$.

Let us now pass the finite $\eta$ value in Eq.~(\ref{ppp}) to the measurements by defining the following POVM elements for Alice:
\begin{align}
\label{povmeta}
M^{\eta}_{a=0|x}&=\eta \bar M_{a=0|x},\nonumber\\
M^{\eta}_{a=1|x}&=\one-\eta \bar M_{a=0|x},
\end{align}
for $x=(1\ldots,N)$. Indeed, with these lossy measurements we have
\begin{equation}
\label{pppp}
p(ab|xy)=\Tr{\left(\rho M^{\eta}_{a|x}\otimes \bar M_{b|y}\right)},
\end{equation}
which gives the same statistics as Eq.~(\ref{ppp}) violating the $N$-setting $I_{NN22}$ inequality for $\eta=1/(N-1)$. However, if we pick any $(N-1)$ measurements from the set defined by the POVM elements~(\ref{povmeta}) above, they turn out to be $(N-1)$-wise jointly measurable for the parameter $\eta=1/(N-1)$. The proof is analogous to the one presented in Appendix~E of Ref~\cite{pauldani}, and is as follows.

Let us consider $n$ lossy two-outcome measurements. We start with arbitrary measurements two outcomes each, $M_{a|x}$, where $a=0,1$ and $x=(1,\ldots,n)$. Then the lossy sets are constructed as follows
\begin{align}
\label{Mlambda}
M^{\eta}_{0|x}&=\eta M_{0|x},\nonumber\\
M^{\eta}_{1|x}&=\one - \eta M_{0|x}.
\end{align}
Clearly, these measurements define valid POVM elements for all $x$. It is proven below that any such set of $n$ measurements is in fact jointly measurable in case of $\eta \leq 1/n$. Let us consider a parent POVM $\{M_\textbf a\}$ with $2^n$ elements, where $\textbf a$ is a length $n$ binary string. Let all the POVM elements $M_\textbf a$ vanish except the ones corresponding to the strings $01\ldots11$, $10\ldots11$, $\ldots$, $11\ldots01$, $11\ldots10$ (that is, when the string contains a single $0$), and $11\ldots11$ (that is, all digits are 1). In these cases, we have the following elements:
\begin{align}
M_{01\ldots11}&=(1/n)M_{0|1},\nonumber\\
M_{10\ldots11}&=(1/n)M_{0|2},\nonumber\\
\vdots\quad&=\quad\vdots\nonumber\\
M_{11\ldots01}&=(1/n)M_{0|n-1},\nonumber\\
M_{11\ldots10}&=(1/n)M_{0|n},\nonumber\\
M_{111\ldots11}&=(1/n)\sum_{x=1}^n M_{1|x}.
\end{align}
If we consider a parent POVM defined by Eq.~(\ref{parent}), we indeed recover the measurements appearing in equation~(\ref{Mlambda}) with $\eta=1/n$. Using this result, we let $n=(N-1)$, and identify any $N-1$ measurements in the set~(\ref{povmeta}) by the parameter $\eta=1/(N-1)$ with the set~(\ref{Mlambda}). This proves that the set of $N$ specific measurements defined by Eq.~(\ref{povmeta}) are $(N-1)$-wise jointly measurable in the case of $\eta\le 1/(N-1)$.

\bigskip

\section{Conclusion}
\label{conclusion}

We investigated the link between Bell nonlocality and incompatibility of measurements and proved that there exists a set of three incompatible dichotomic qubit measurements which never give rise to Bell nonlocality. We recall that this is the simplest situation in which the two notions may differ, since for a pair of dichotomic measurements it has been proved by Wolf et al.~\cite{wolf} that measurement incompatibility entails violation of Bell inequalities. Recently, the case of more than two dichotomic measurements have been addressed. Importantly, Quintino et al.~\cite{q16} constructed a LHV model for a set of incompatible qubit measurements. The present study can be considered as a generalization of Ref.~\cite{q16} in different aspects: On one hand, Bob's two outcome settings have been generalized to measurement settings with arbitrary outcomes. On the other, Alice's set of measurements could be decreased from an infinite number to the minimum number of three settings. Note also a more recent work~\cite{q17} obtaining related results.

Moving away from the bipartite case, we can ask the following question. Does there exist a set of incompatible measurements such that if Alice measures this set independently of the set of measurements chosen by Bob and Charlie and the three-party state shared by them, the resulting statistics is not genuinely tripartite nonlocal (in the sense of not able to violate any Svetlichny-type inequality~\cite{svet1,svet2,svet3})? This question can be considered as a generalization of the two-party case to more parties.

Finally, we presented a set of $N$ suitably chosen measurements in dimension $N$, which are $(N-1)$-wise jointly measurable, such that they provide a Bell violation. It remains an open problem if such a set of $N$ measurements can be found in the case of minimal dimension 2.

\section{Acknowledgements} We thank Antonio Ac\'in and Marco T\'ulio Quintino for helpful discussions. We acknowledge the support of the National Research, Development and Innovation Office NKFIH (Grant Nos. K111734 and KH125096).

\appendix

\section{Real-valued unitaries}
\label{appnew}

We prove that one can choose $U_A=O(\varphi)$ and $U_B=\one$ in the state~(\ref{psi}) without the loss of generality, where $O(\varphi)$ is the planar rotation~(\ref{Orot}) and Bob's qubit measurements $\{M_{b|y}\}$ are real valued.

%This also entails that Bob's real-valued measurements have at most three outcomes (that is, $b\in\{0,1,2\})$.

Suppose that the distribution $p(ab|xy)$ in Eq.~(\ref{stat}) is local for all $\{M_{b|y}\}$ real valued, however, it lies outside the local set (i.e. nonlocal) for $\{M_{b|y}\}$ complex valued. Let's denote this nonlocal distribution by $p'(ab|xy)$. We next show that this situation cannot occur. Hence this is a proof by contradiction.

Since the LHV set~(\ref{LHV}) is convex, the nonlocal distribution $p'(ab|xy)$ implies that there must exist a hyperplane with
associated (real-valued) Bell coefficients $c_{ab|xy}$, such that
\begin{equation}
\label{bigger} \beta \equiv
\sum_{a,b,x,y}c_{ab|xy}p'(ab|xy)>\max\sum_{a,b,x,y}c_{ab|xy}p(ab|xy),
\end{equation}
where maximization is over all $p(ab|xy)$ within the LHV set. However, as we will show the value of $\beta$ in Eq.~(\ref{bigger}) can also be attained with $U_A=O(\varphi)$ and $U_B=\one$ and real valued qubit measurements $M_{b|x}$
for Bob. Hence, there exists some nonlocal distribution $p(ab|xy)$ where the set $\{M_{b|y}\}$ is real-valued, which is a contradiction.

We now show that $\beta$ can be attained using $U_A=O(\varphi)$ and $U_B=\one$ and a real valued set $\{M_{b|y}\}$. To this end, let us denote
\begin{equation}
\label{sigmaby}
\sigma_{b|y}=\Tr_B{(\rho \one\otimes M_{b|y})},
\end{equation}
and let
\begin{equation}
\label{Fby}
F_{b|y} = \sum_{a,x}c_{ab|xy}M_{a|x}^{\eta^*}.
\end{equation}
With these, we have $\beta=\sum_{b,y}\Tr{(F_{b|y}\sigma_{b|y})}$. Since $M_{a|x}^{\eta^*}$ is real valued, $F_{b|y}$ are symmetric matrices. Then, by redefining $\sigma_{b|y}$ as $(\sigma_{b|y}+\sigma^*_{b|y})/2$, we get a real-valued assemblage, which provides the same $\beta$ value in Eq.~(\ref{bigger}). Due to the GHJW construction \cite{ghjw1,ghjw2}, any such real valued no-signalling qubit assemblage $\{\sigma_{b|y}\}$ has a quantum realization with a state $\rho=\ket{\psi}\bra{\psi}$ in the form
\begin{equation}
\label{schmidt}
\ket{\psi}=\sum_{i=0,1}{\sqrt{\lambda_i}(O_A(\varphi)\ket{i})\ket{i}},
\end{equation}
where $\lambda_i$ are positive Schmidt coefficients and $O_A$ is the orthogonal qubit matrix defined by~(\ref{Orot}). These can be obtained through the diagonalization $\sigma_A = \sum_b{\sigma_{b|y}}=O_A(\varphi)\sum_i{\lambda_i\ket{i}\bra{i}}O_A(\varphi)$. On the other hand, Bob's measurements $M_{b|y}$ can be written in the form
\begin{equation}
\label{bobpovm}
M_{b|y}=\sum_{i=0,1}\sum_{j=0,1}\frac{1}{\sqrt{\lambda_i\lambda_j}\ket{i}\bra{j}\sigma_{b|y}\ket{i}\bra{j}},
\end{equation}
which define valid real-valued qubit POVM elements (as they are readily positive and sum up to the identity).

\section{Computation of $\eta(\theta,\varphi_j)$}

\label{appB}

We have the special case of equation~(\ref{pvsigma}), where $\varphi_j$ is fixed:
\begin{equation}
\label{pvsigmaj}
v\rho(\theta,\varphi_j)+(1-v)\frac{\one}{2}\otimes\Tr_A{\rho(\theta,\varphi_j)}=p \rho(\theta_i,\varphi_j) + (1-p)\sigma
\end{equation}
Then we have $\eta(\theta,\varphi_j)=v\eta(\theta_i,\varphi_j)$. First let us observe that we can rotate Alice's system by an angle $-\varphi_j$, such that we get the same $v$ in the un-rotated system. Then it is enough to determine $v$ and $p$ in the decomposition~(\ref{pvsigmaj}) when $\varphi_j=0$.

Our goal is to get a good lower bound on $v$ in the function of $\delta\theta=\theta_i-\theta>0$.
Following similar steps as in the derivation carried out in Ref.~\cite{q16}, that is constraining that $\sigma$ is a diagonal matrix in Eq.~(\ref{pvsigmaj}), and demanding the positivity of the diagonal elements of $\sigma$, we get the following upper bound formulas for $v$:
\begin{align}
\label{sigmadiag}
\sigma_{00,00}&\ge 0 \rightarrow v\le (\tan\theta\cot\theta_i(1+\eta)-\eta)^{-1},\nonumber\\
\sigma_{01,01}&\ge 0 \rightarrow v\le (\cot\theta\tan\theta_i(1-\eta)+\eta)^{-1},\nonumber\\
\sigma_{10,10}&\ge 0 \rightarrow v\le (\tan\theta\cot\theta_i(1-\eta)+\eta)^{-1},\nonumber\\
\sigma_{11,11}&\ge 0 \rightarrow v\le (\cot\theta\tan\theta_i(1+\eta)-\eta)^{-1},
\end{align}
where $\eta$ above is expressed by the angles $(\theta_i,\varphi_j)$ and we also assume that $\theta\le\theta_i$. It turns out that the smallest value corresponds to the last line, which is the most constraining, hence we can take
\begin{equation}
\label{etaformula}
\eta(\theta,\varphi_j) = \frac{\eta(\theta_i,\varphi_j)}{\cot\theta\tan\theta_i(1+\eta(\theta_i,\varphi_j))-\eta(\theta_i,\varphi_j)}.
\end{equation}
It is noted that in the other case of $\theta\ge\theta_i$, the most constraining relation in Eqs.~(\ref{sigmadiag}) corresponds to the first line.

\section{Computation of $\eta(\theta_i,\varphi)$}

\label{appC}

We have the special case of equation~(\ref{pvsigma}), when $\theta_i$ is fixed:
\begin{equation}
\label{pvsigmai}
v\rho(\theta_i,\varphi)+(1-v)\frac{\one}{2}\otimes\Tr_A{\rho(\theta_i,\varphi)}=p \rho(\theta_i,\varphi_j) + (1-p)\sigma.
\end{equation}
Then we have $\eta(\theta_i,\varphi)=v\eta(\theta_i,\varphi_j)$. We can rotate Alice's system by an angle $-\varphi_j$, such that we get the same $v$ in the un-rotated system. Then it is enough to determine $v$ and $p$ in the decomposition
\begin{equation}
\label{p0vsigmai}
v\rho(\theta_i,\delta\varphi)+(1-v)\frac{\one}{2}\otimes\Tr_A{\rho(\theta_i,\delta\varphi)}=p \rho(\theta_i,0) + (1-p)\sigma.
\end{equation}
We wish to get a good lower bound on $v$ in the function of $\delta\varphi=\varphi-\varphi_j>0$ for fixed $\theta_i$.

To this end, we prove that we can take $p=v$ in Eq.~(\ref{p0vsigmai}) above, where $v$ is given by
\begin{equation}
\label{vformula}
v(\delta\varphi) = \frac{1-2\sqrt 2\sqrt{1-\cos(2\delta\varphi)}}{8\cos(2\delta\varphi)-7},
\end{equation}
which results in $\sigma$ separable. Indeed, if we rearrange equation~(\ref{p0vsigmai}) for $\sigma$, it will take the form
\begin{equation}
\label{sigmasep}
\sigma = \frac{v}{1-v}\left(\rho(\theta_i,\delta\varphi)-\rho(\theta_i,0)\right)+\frac{\one}{2}\otimes\Tr_A{\rho(\theta_i,\delta\varphi)}.
\end{equation}
If we insert $v$ from (\ref{vformula}) into (\ref{sigmasep}), one can see that $\sigma$ is a valid two-qubit separable state. This can be checked by first noting that $\text{PT}(\sigma)=\sigma$ (for arbitrary $v$), where PT denotes partial transposition~\cite{peres,horoppt} with respect to system B. On the other hand, $\sigma$ is a valid density matrix. Readily, $\Tr{\sigma}=1$ and all its eigenvalues turn out to be positive
\begin{align}
\lambda_1 &=0,\nonumber\\
\lambda_2 &=\frac{1}{4},\nonumber\\
\lambda_{3,4} &=\frac{3\pm\sqrt{5+4\cos(4\theta_i)}}{8}
\end{align}
for any $\theta_i$. Then the relation $\eta(\theta_i,\varphi)=v(\varphi-\varphi_j)\eta(\theta_i,\varphi_j)$ follows, where $v$ is given by Eq.~(\ref{vformula}).


\vfill

\begin{thebibliography}{99}

\bibitem{bell}
J.S. Bell, On the Einstein-Podolsky-Rosen paradox, \emph{Physics} {\bfseries 1}, 195--200 (1964).

\bibitem{brunnerreview}
N. Brunner, D. Cavalcanti, S. Pironio, V. Scarani, and S. Wehner, Bell nonlocality, \emph{Rev. Mod. Phys.} {\bfseries 86}, 419--478 (2014).

\bibitem{werner}
R.F. Werner, Quantum states with Einstein-Podolsky-Rosen correlations
  admitting a hidden-variable model, \emph{Phys. Rev. A}
  {\bfseries 40}, 4277--4281 (1989).

\bibitem{barrett} J. Barrett, Nonsequential positive-operator-valued measurements on entangled mixed states do not always violate a Bell inequality, \emph{Phys. Rev. A} {\bfseries 65}, 042302 (2002).

\bibitem{betterhirsch} F. Hirsch, M. T. Quintino, T. V\'ertesi, M. Navascu\'es, N. Brunner, Better local hidden variable models for two-qubit Werner states and an upper bound on the Grothendieck constant $KG(3)$, \emph{Quantum} {\bf 1}, 3 (2017).

\bibitem{simpovm}
M. Oszmaniec, L. Guerini, P. Wittek, A. Ac\'in, Simulating Positive-Operator-Valued Measures with Projective Measurements, \emph{Phys. Rev. Lett.} {\bf 119}, 190501 (2017).

\bibitem{khalfin}
L.A. Khalfin, B.S. Tsirelson, Quantum and quasi-classical analogs of Bell inequalities, \emph{Symposium on the Foundations of Modern Physics} 441--460 (1985).

\bibitem{wolf} M.M. Wolf, D. Perez-Garcia, and C. Fernandez, Measurements
  Incompatible in Quantum Theory Cannot Be Measured Jointly in Any Other
  No-Signaling Theory, \emph{Phys. Rev. Lett.} {\bfseries 103}, 230402 (2009).

\bibitem{uola14}
R. Uola, T. Moroder, and O. G{\"u}hne, Joint Measurability of Generalized Measurements Implies Classicality, \emph{Phys. Rev.
  Lett.} {\bfseries 113}, 160403 (2014).

\bibitem{q14}
M.T. Quintino, T. V{\'e}rtesi, and N. Brunner, Joint Measurability, Einstein-Podolsky-Rosen Steering, and Bell Nonlocality, \emph{Phys. Rev. Lett.} {\bfseries 113}, 160402 (2014).

\bibitem{q16}
M.T. Quintino, J. Bowles, F. Hirsch, N. Brunner, Incompatible quantum measurements admitting a local hidden variable model, \emph{Phys. Rev. A} {\bfseries 93}, 052115 (2016).

\bibitem{EPR1} A. Einstein, B. Podolsky, and N. Rosen, Can quantum-mechanical description of physical reality be considered complete?, \emph{Phys. Rev.} {\bf 47}, 777–780, (1935).

\bibitem{EPR2} H.~M. Wiseman, S.~J. Jones, and A.~C. Doherty, Steering, entanglement, nonlocality, and the Einstein-Podolsky-Rosen paradox, \emph{Phys. Rev. Lett.} {\bf 98}, 2, (2007).

\bibitem{busch}
P. Busch, Unsharp reality and joint measurements for spin observables, \emph{Phys. Rev. D}, {\bfseries 33}, 2253 (1986).

\bibitem{buschbook}
P. Busch, P. Lahti, and P. Mittelstaedt, {\em The Quantum Theory of
  Measurement}. \newblock Lecture Notes in Physics Monographs Vol. 2, Springer 1996, pp 25-90.

\bibitem{liang}
Y.C. Liang, R.W. Spekkens, and H.M. Wiseman, Specker's parable of
  the overprotective seer: A road to contextuality, nonlocality and
  complementarity, \emph{Phys. Rep.} {\bfseries 506}, 1--39 (2011).

\bibitem{hrs} T. Heinosaari, D. Reitzner, and P. Stano, Notes on Joint
Measurability of Quantum Observables, \emph{Foundations of
Physics} {\bfseries 38}, 1133–-1147 (2008).

\bibitem{uola16}
R. Uola, K. Luoma, T. Moroder, T. Heinosaari, Adaptive strategy for joint measurements, \emph{Phys. Rev. A} {\bf 94}, 022109 (2016).

\bibitem{dariano05}
G. Mauro D'Ariano, P. Lo Presti, and P. Perinotti, Classical
  randomness in quantum measurements, \emph{J. Phys. A: Math. Gen.} {\bfseries 38}, 5979--5991 (2005).

\bibitem{hirsch16}
F. Hirsch, M.T. Quintino, J. Bowles, T. V\'ertesi, N. Brunner, Entanglement without hidden nonlocality, \emph{New J. Phys.} {\bf 18}, 113019 (2016)

\bibitem{finitesimulate}
J. Bowles, F. Hirsch, M. T. Quintino, and N. Brunner, Local Hidden Variable Models for Entangled Quantum States Using Finite Shared Randomness, \emph{Phys. Rev. Lett.} {\bf 114}, 120401 (2015).

\bibitem{pironio}
S. Pironio, All CHSH polytopes, \emph{J. Phys. A: Math. Theor.} {\bf 47}, 424020 (2014).

\bibitem{chsh}
J.F. Clauser, M.A. Horne, A. Shimony, and R.A. Holt, Proposed Experiment to Test Local Hidden-Variable Theories, \emph{Phys. Rev. Lett.} {\bfseries 23}, 880--884 (1969).

\bibitem{horo}
R. Horodecki, P. Horodecki, and M. Horodecki, Violating Bell inequality by mixed spin-$1/2$ states: necessary and sufficient condition, \emph{Phys. Lett. A}, {\bf 200}, 340 (1995).


\bibitem{alg1} F. Hirsch, M.T. Quintino, T. V\'ertesi, M.F. Pusey, and N. Brunner, Algorithmic construction of local hidden variable models for entangled quantum states, \emph{Phys. Rev. Lett.} {\bf 117}, 190402 (2016).

\bibitem{cdd}
K. Fukuda, cdd program, 2003. \url{https://www.inf.ethz.ch/personal/fukudak/cdd_home/}.

\bibitem{alg2} D. Cavalcanti, L. Guerini, R. Rabelo, and  P. Skrzypczyk, General method for constructing local-hidden-variable models for entangled quantum states, \emph{Phys. Rev. Lett.} {\bf 117}, 190401 (2016).

\bibitem{mosek}
MOSEK ApS, 2016, The MOSEK optimization toolbox for MATLAB manual. Version 7.1 (Revision 28).
 \url{http://docs.mosek.com/7.1/toolbox/index.html}.

\bibitem{amoeba}
W. Press, S. Teukolsky, W. Vetterling, and B. Flannery, \emph{Numerical Recipes: The Art of Scientific Computing}, (Cambridge University Press, New York, 2007).

\bibitem{I3322}
D. Collins and N. Gisin, A relevant two qubit Bell inequality inequivalent to the CHSH inequality, \emph{J. Phys. A} {\bf 37}, 1775 (2004).

\bibitem{closing} T. V\'ertesi, S. Pironio, and N. Brunner, Closing the Detection Loophole in Bell Experiments Using Qudits, \emph{Phys. Rev. Lett.} {\bf 104}, 060401 (2010).

\bibitem{pauldani}
P. Skrzypczyk and D. Cavalcanti, Loss-tolerant Einstein-Podolsky-Rosen steering for arbitrary-dimensional states: Joint measurability and unbounded violations under losses, \emph{Phys. Rev. A} {\bf 92}, 022354 (2015).

\bibitem{ghjw1}
N. Gisin, Stochastic quantum dynamics and relativity, \emph{Helv. Phys. Acta} {\bf 62}, 363 (1989).

\bibitem{ghjw2}
L.P. Hughston, R. Jozsa, and W.K. Wootters, A complete classification of quantum ensembles having a given density matrix, \emph{Phys. Lett. A} {\bf 183}, 14 (1993).

\bibitem{peres}
A. Peres, Separability Criterion for Density Matrices, \emph{Phys. Rev. Lett.} {\bfseries 77}, 1413--1415 (1996).

\bibitem{horoppt}
M. Horodecki, P. Horodecki, and R. Horodecki, Separability of mixed states: necessary and sufficient conditions, \emph{Physics Letters A}, {\bfseries 223}, 1--8 (1996).

\bibitem{q17}
F. Hirsch, M. T. Quintino, N. Brunner, Quantum measurement incompatibility does not imply Bell nonlocality, arXiv:1707.06960 (2017).

\bibitem{svet1}
G. Svetlichny, Distinguishing three-body from two-body nonseparability by a Bell-type inequality, \emph{Phys. Rev. D} {\bf 35}, 3066 (1987).

\bibitem{svet2} R. Gallego, L. E. W\"urflinger, A. Ac\'in, and M. Navascu\'es, Operational Framework for Nonlocality, \emph{Phys. Rev. Lett.} {\bf 109}, 070401 (2012).

\bibitem{svet3} R. Chaves, D. Cavalcanti, and L. Aolita, Causal hierarchy of multipartite Bell nonlocality, \emph{Quantum} {\bf 1}, 23 (2017).

\end{thebibliography}
\end{document}